\documentclass[aps,pra,reprint]{revtex4-1}
\usepackage{amsfonts, amssymb, amsmath,graphicx}
\usepackage{subfigure}
\usepackage{multirow}
\usepackage{dcolumn}
\usepackage{bm}
\usepackage{color}

\begin{document}

\newcommand{\tom}[1]{{\color{red} #1}}

\title{Artificial magnetic field for synthetic quantum matter without dynamical modulation}

\author{Tomoki Ozawa}
\affiliation{Advanced Institute for Materials Research, Tohoku University, Sendai 980-8577, Japan}
\affiliation{RIKEN Interdisciplinary Theoretical and Mathematical Sciences Program, Wako 351-0198, Japan}%

\date{\today}

\begin{abstract}
We propose an all-static method to realize an artificial magnetic field for charge neutral particles without introducing any time modulation. Our proposal consists of one-dimensional tubes subject to harmonic trapping potentials with shifted centers. We show that this setup realizes an artificial magnetic field in a hybrid real-momentum space. We discuss how characteristic features of particles in a magnetic field, such as chiral edge states and the quantized Hall response, can be observed in this setup. We find that the mean-field ground state of bosons in this setup in the presence of long-range interactions in physical real space can have quantized vortices in the hybrid real-momentum space; such a state with vortices exhibits a supersolid structure in the physical real space. Our method can be applied to a variety of synthetic quantum matter, including ultracold atomic gases, coupled photonic cavities, coupled waveguides, and exciton-polariton lattices. 
\end{abstract}

\maketitle

\noindent \textit{Introduction}
Quantum simulation has been a central theme in the research of synthetic quantum matter based on atomic, molecular, and optical systems, such as ultracold atomic gases, trapped ions, polaritons, and photonics~\cite{Bloch:2012, Blatt:2012, Guzik:2012, Carusotto:2013, Georgescu:2014, Bloch:2018, Boulier:2020}.
In particular, simulation of topological phenomena using synthetic quantum matter has attracted a considerable attention in the recent years~\cite{Hasan:2010, Qi:2011, Chiu:2018, Armitage:2018, Cooper:2019, Ozawa:2019}.
A prototypical example of topological phenomena is the quantum Hall effect, which was first found in a two-dimensional electron gas under a magnetic field~\cite{Klitzing:1980, Tsui:1982, Yoshioka:Book}.
To simulate the quantum Hall effect and other topological phenomena, one often needs to simulate physics of charged particles in a magnetic field, which is not straightforward because atoms in ultracold gases and photons in cavities and waveguides are charge-neutral.

Various methods have been employed to simulate effects of a magnetic field, namely to realize an {\it artificial} magnetic field~\cite{Aidelsburger:2018}.
In ultracold atomic gases, the earliest example is to rotate the system, exploiting the similarity between the Coriolis force and the Lorentz force~\cite{Madison:2000, Cooper:2008, Fetter:2009}. Other examples include Floquet engineering~\cite{GoldmanDalibard:2014}, light-induced gauge potential~\cite{Goldman:2014}, and synthetic dimensions~\cite{Ozawa:2019SD}. Most of the existing proposals involve adding fast time modulation and/or exploiting different energy and time scales present in the system to explore adiabatic physics in a low energy subspace. These methods often suffer from instability and heating as a result of time modulation and spontaneous emissions. A notable exception is the method using spinor condensates in a quadrupole magnetic field where spatially dependent spin texture results in an artificial magnetic field~\cite{Ho:1996,Pietila:2009,Choi:2013,Ray:2014}. The situation is similar in photonics and optics. For light close to optical frequencies, magneto-optical effects are generally weak and thus, similar to ultracold atomics gases, one needs to realize an artificial magnetic field to explore the physics of quantum Hall effects~\cite{Rechtsman:2013, Yuan:2018, Lustig:2019}. Static realizations of topological models in photons are in the class of quantum spin-Hall insulators where two degrees of freedom feel opposite magnetic fields and thus topological protection relies upon decoupling of different degrees of freedom~\cite{Hafezi:2013, Schine:2016}. 

\begin{figure}[t]
  \centering
  \includegraphics[width=0.95\columnwidth]{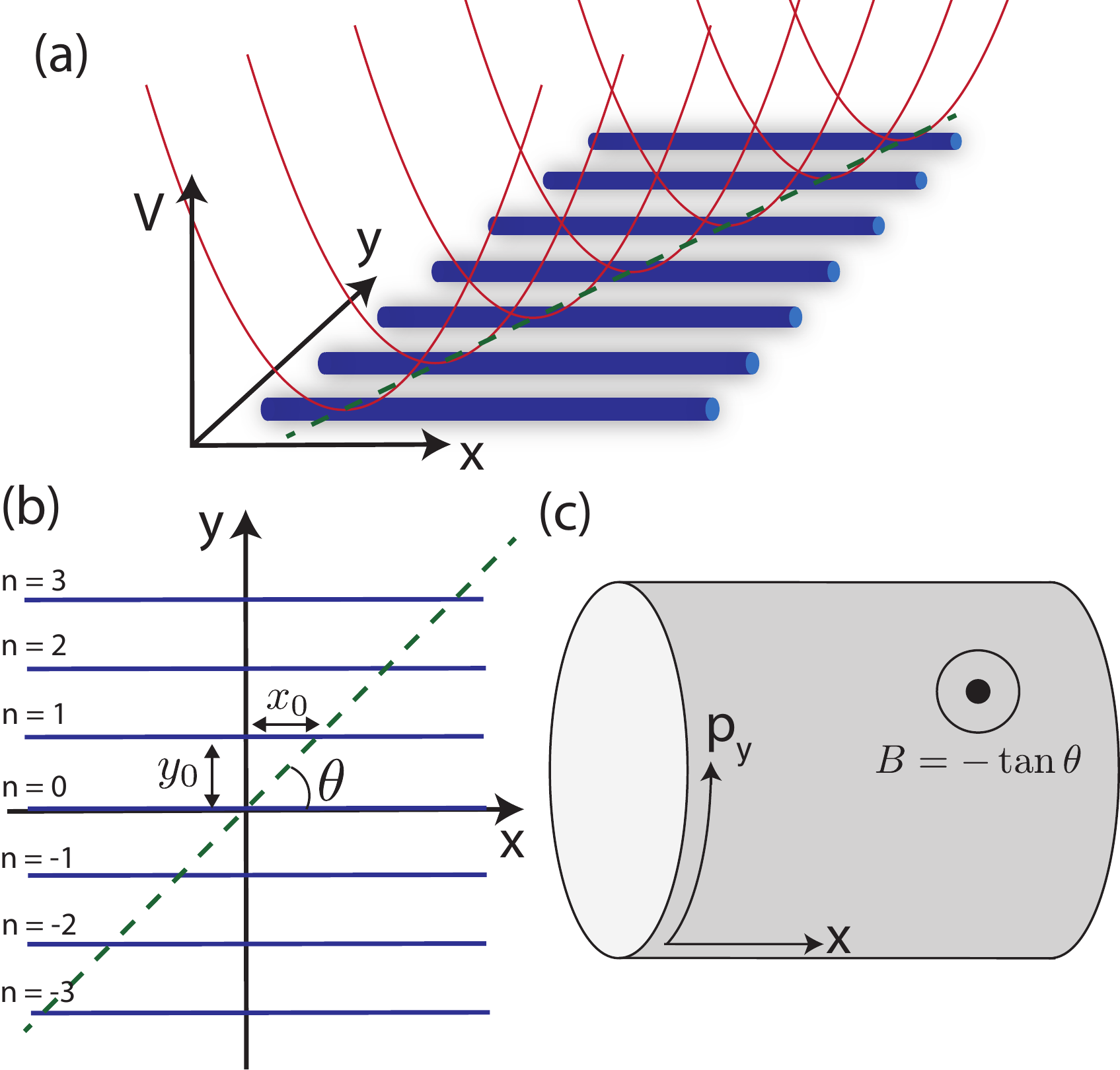}
  \caption{Illustration of the scheme. (a) One-dimensional tubes elongated in the $x$ direction are aligned along the $y$ direction. Harmonic trapping potentials, depicted by parabolas, are applied to the tubes. The trapping centers are on the dashed line, which is tilted with respect to the $x$ axis by an angle $\theta$. (b) Setup viewed from top. Tubes are located at $y = n y_0$, where $n$ is an integer. The trapping centers are on the dashed line $y = x \tan \theta$. (c) In the $x$-$p_y$ plane, an artificial magnetic field of strength $B = -\tan \theta$ is realized. We have the geometry of a cylinder, where the $p_y$ direction is periodic with periodicity $2\pi/y_0$.
  }\label{fig:setup}
\end{figure}

In this paper, we propose a completely static way to realize an artificial magnetic field regardless of the spin of the particles, which can be applied to a variety of synthetic quantum matter.
Our proposal is based on a well-known fact that the Hamiltonian of a charged quantum particle in two dimensions under a uniform magnetic field is equivalent to a set of harmonic oscillators with shifted centers~\cite{LL:QM}. 
This fact implies, conversely, that a set of harmonic oscillators with shifted centers can be viewed as a charged particle in a uniform magnetic field.
We pursue this analogy further and consider a set of one-dimensional systems elongated along the $x$ direction and align these one-dimensional tubes in the $y$-direction and place them under harmonic trapping potentials with shifted centers (Fig.~\ref{fig:setup}). We find that the Hamiltonian of a charged particle in a magnetic field is realized in the $x$-$p_y$ plane where $p_y$ is the momentum along the $y$ direction. Even though the proposal looks deceptively simple, as we discuss in detail, we can still observe most of the phenomena characteristic of charged particles in a magnetic field, such as the chiral edge states and quantized Hall response. We note that there is no dynamical component in the proposal, and, in particular, our proposal does not break the time-reversal symmetry in the physical $x$-$y$ plane. Our proposal is thus expected to be more stable and free from heating compared with existing proposals which rely on dynamical modulations. 

We also explore the effect of interparticle interactions in the artificial magnetic field. Weakly interacting bosons under a magnetic field form vortices~\cite{Madison:2000, Cooper:2008, Fetter:2009, Spielman:2009, Lin:2009}. However, a contact interaction, common in ultracold atomic gases and photons with optical nonlinearity, translates into extremely long-range interaction in the $p_y$ direction, and we find that vortices do not form under ordinary contact interactions. On the other hand, by including nearest-neighbor interaction in $y$ direction we find that vortices can form in the $x$-$p_y$ plane. Such a vortex structure corresponds to supersolid structure in the physical $x$-$y$ plane, where the phase coherence of a condensate is maintained and the discrete translational symmetry is broken. 
Finally, we discuss possible experimental platforms where the proposal can be implemented.

\section{Setup}
We consider a set of uncoupled one-dimensional tubes elongated in the $x$ direction and aligned in the $y$ direction, where each tube is subject to a harmonic trapping potential. The center of the trapping potential shifts between tubes by a constant amount $x_0$. We assume that tubes are separated by a constant amount $y_0$ in the $y$ direction. See Fig.~\ref{fig:setup} for an illustration. We use the second-quantized formalism, where $\hat{\psi}^\dagger_n(x)$ and $\hat{\psi}_n(x)$ are the creation and annihilation operators, respectively, of a particle at coordinate $(x, n y_0)$, where $n$ indicates that the particle is in the $n$th tube.
The single-particle Hamiltonian is
\begin{align}
	\hat{H} =& \sum_n \int dx  \left[ \frac{1}{2m} \left( \partial_x \hat{\psi}_n^\dagger (x)\right) \left( \partial_x \hat{\psi}_n (x)\right)
	\right. \notag \\
	&+\left.
	\frac{1}{2}m\omega^2 (x - nx_0)^2 \hat{\psi}_n^\dagger (x) \hat{\psi}_n (x)
	\right], \label{startingpoint}
\end{align}
Note that there is no motion (coupling) between different tubes in the $y$ direction.
By performing Fourier transformation in the $y$ direction,
\begin{align}
	\hat{\psi}_n (x) = \sqrt{\frac{y_0}{2\pi}}\int_{-\pi/y_0}^{\pi/y_0} dp_y e^{iny_0 p_y} \hat{\psi}(x,p_y),
\end{align}
the Hamiltonian can be rewritten as
\begin{align}
	\hat{H}
	=&
	\int dx \int_{-\pi/y_0}^{\pi/y_0} d p_y
	\left[
	\frac{1}{2m} \left( \partial_x \hat{\psi}^\dagger (x,p_y)\right)\left( \partial_x \hat{\psi} (x,p_y)\right)
	\right. \notag \\
	&+\left.
	\frac{1}{2} \frac{m \omega^2}{\tan^2 \theta}
	\left\{ (i\partial_{p_y} + x \tan \theta )\hat{\psi}^\dagger (x,p_y) \right\}
	\right.
	\notag \\
	&\hspace{2cm}\left.
	\left\{ (-i\partial_{p_y} + x \tan \theta )\hat{\psi} (x,p_y) \right\}
	\right].
\end{align}
In the first-quantized form, this Hamiltonian is 
\begin{align}
	H = -\frac{1}{2m} \partial_x^2 + \frac{m\omega^2}{2\tan^2 \theta}\left( -i\partial_{p_y} + x \tan \theta \right)^2
\end{align}
This is nothing but the Hamiltonian of a charged particle in the $x$-$p_y$ plane under a magnetic vector potential $\mathbf{A}(x,p_y) = (0, -x \tan \theta)$ in the Landau gauge, corresponding to the artificial magnetic field $B = -\tan \theta$ penetrating the plane. (We have set the charge $e$, speed of light $c$, and Plank constant $\hbar$ to be unity.)
We have chosen to use $-i\partial_{p_y} = -y$ as the ``momentum" operator along the $p_y$ direction, regarding $p_y$ as the synthetic ``position."
The mass is anisotropic; the mass in the $x$ direction is the original mass $m_x = m$, whereas the mass in the $p_y$ direction is $m_{p_y} = \tan^2 \theta/m\omega^2$. This anisotropy can be absorbed by rescaling units of length in two directions and does not pose any problem in the rest of the discussion.
Another noticeable feature here is that the $x$-$p_y$ plane is a cylinder, open along the $x$ direction and periodic along the $p_y$ direction.

Let us comment on the time-reversal symmetry in our system. Our original Hamiltonian (\ref{startingpoint}) does not break the physical time-reversal symmetry. For a single component (spinless) system, the time-reversal operation flips the sign of the momentum $\mathbf{p} \to -\mathbf{p}$. Since we want to look at the physics related to the artificial magnetic field in the $x$-$p_y$ plane, the relevant effective time-reversal symmetry is now to flip their reciprocal variables $\mathcal{T}: (p_x, -y) \to (-p_x, y)$; the effective time-reversal symmetry is the simultaneous operation of flipping of $p_x$ and the parity in $y$. Our Hamiltonian does break the parity in the $y$ direction due to the shifted harmonic trapping potentials, which enables us to realize the artificial magnetic field in the $x$-$p_y$ plane without breaking the physical time-reversal symmetry.

Our method is related to the idea of the topological charge pumping, where a momentum is replaced by an external parameter; by sweeping the parameter one obtains topological properties of the original Hamiltonian~\cite{Thouless:1983, Nakajima:2016, Lohse:2016}. Our method, however, is crucially different from charge pumping in that we replace a momentum $p_y$ by a physical real coordinate $-y$, and in this way we succeed in recovering the full dynamics in all dimensions, whereas no dynamics occurs in charge pumping in the direction replaced by an external parameter.

\begin{figure}[t]
  \centering
  \includegraphics[width=\columnwidth]{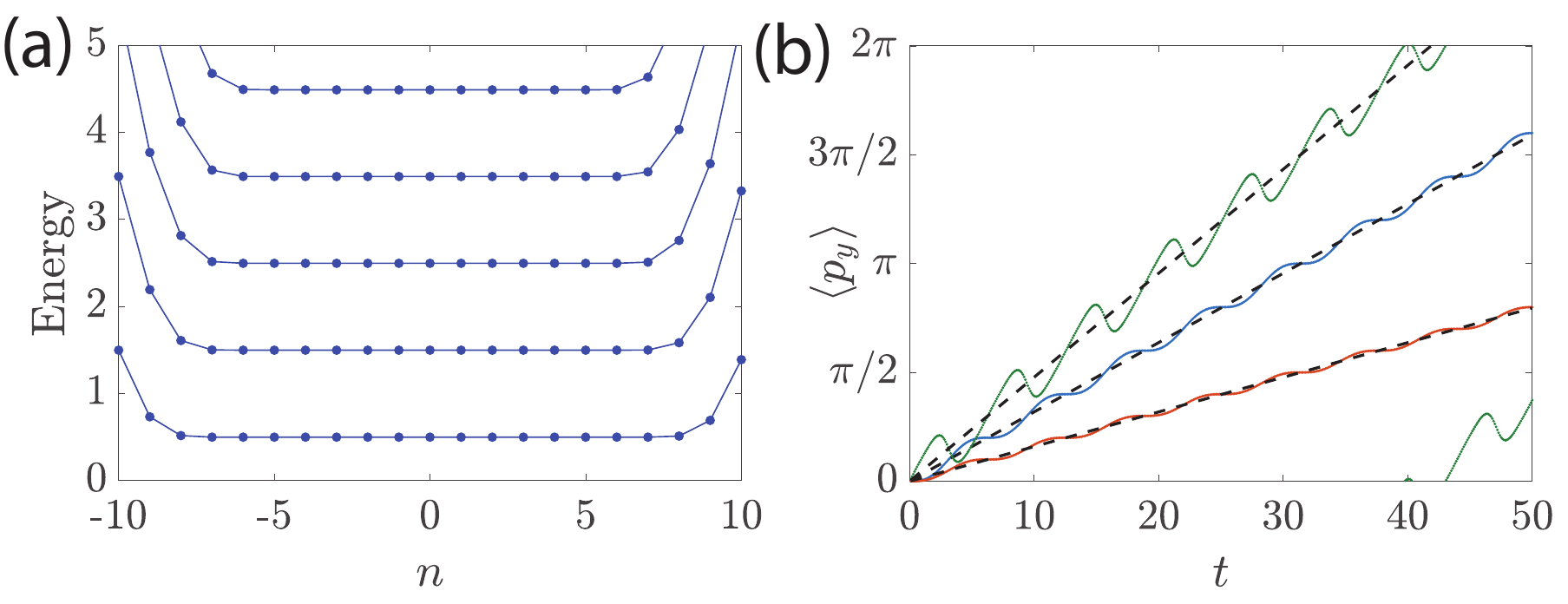}
  \caption{Energy level and the Hall response. (a) The energy level of each tube at position $n$, where length of each tube is $20 a_\mathrm{osc}$ and we employ an open boundary condition. We choose $\tan \theta = 1$, and the energy is in units of the cyclotron frequency $\omega$. Solid lines are to guide the eye. (b) The center-of-mass motion of wave packets when a synthetic electric field $E_x = 0.1 \omega/a_{\mathrm{osc}}$ is applied. Three solid lines show the time evolution of the center of mass in the $p_y$ direction for $\tan \theta = 2/3$ (top, green), $\tan \theta = 1$ (middle, blue), and $\tan \theta = 1/2$ (bottom, red). The dashed lines going through the solid lines are theoretical predictions from Eq.~(\ref{halleffect}) with $\mathcal{C} = 1$. The units of time are $1/\omega$, and the units of $p_y$ are $1/y_0$.
  }\label{fig:energynado}
\end{figure}

\section{Chiral edge states}
We now analyze how characteristic features of charged particles in a magnetic field can be observed in our setup in the $x$-$p_y$ plane. We first point out that the cyclotron frequency of the artificial magnetic field is $\omega$, as expected. This can be derived from the definition of the cyclotron frequency $B/\tilde{m}$; as the mass $\tilde{m}$, we need to use a geometric mean of anisotropic effective mass in two directions $\tilde{m} = \sqrt{m_x m_{p_y}}$. Using $B = -\tan \theta$, we then obtain $\omega = |B|/\tilde{m}$. This observation implies the well-known fact that the energy level of harmonic oscillators in the $x$-$y$ plane, which is $\omega (l + 1/2)$, where $l = 0, 1, 2, \cdots$ is a nonnegative integer, is nothing but the Landau-level energy spectrum. High degeneracy of Landau levels come from the many uncoupled tubes present in the system, which all have the same energy spectrum. 

In the presence of edges, charged particles in two dimensions under a magnetic field should have chiral edge states. In Fig.~\ref{fig:energynado}(a), we plot the energy level of the system in the presence of sharp edges in the $x$ direction as a function of $n$, which is the reciprocal variable of $p_y$. We consider 21 tubes and the length of each tube in the $x$ direction is $20 a_{\mathrm{osc}}$, where $a_\mathrm{osc} \equiv 1/\sqrt{m\omega}$ is the oscillator length. We can see that at large values of $|n|$, the dispersion goes upward; these states are localized at the edges and they propagate in one direction.
The existence of chiral edge states can be probed from the dynamics of wave packets localized at the edges as we show below.
Since there is no coupling between different tubes, the population of particles in each tube does not change; in real space, the propagation of chiral edge states in the $p_y$ direction appears as an appropriate evolution of the relative phases in each tube, which can be probed, for example, in ultracold atomic gases through time-of-flight imaging. 

\begin{figure*}[htbp]
  \centering
  \includegraphics[width=1.8\columnwidth]{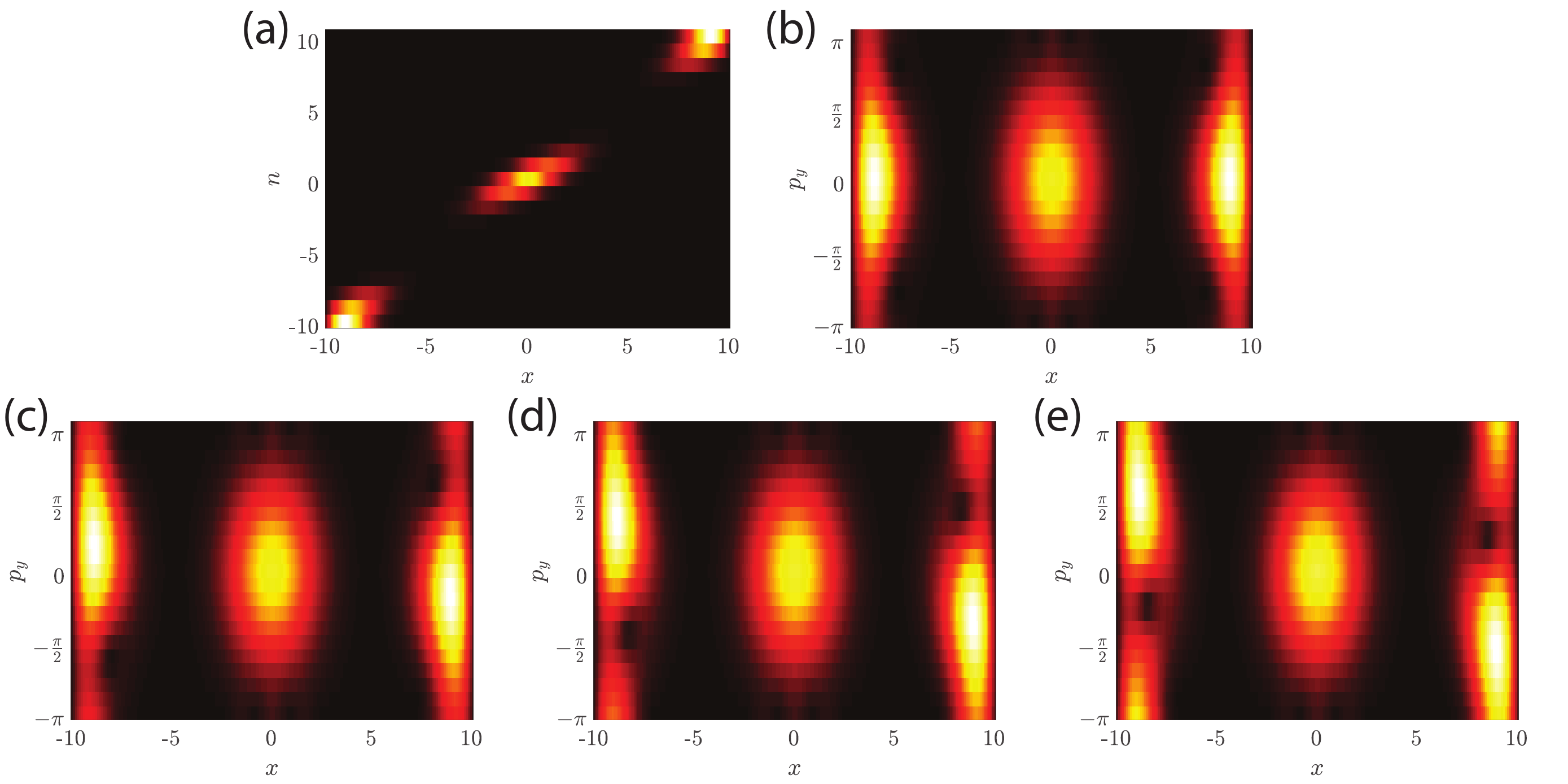}
  \caption{Dynamics of a wave packet and chiral edge modes. The units of length in the $x$ direction are $a_\mathrm{osc}$, and the units of $p_y$ are $1/y_0$. (a) The absolute value of the initial wave function is plotted in the $x$-$y$ plane, where each $n$ represents a tube located at the position $y = n y_0$. (b) The initial state in the $x$-$p_y$ plane. The lower panel shows the wave function in the $x$-$p_y$ plane after time evolution for duration (c) $t = 1/\omega$, (d) $t = 2\omega$, and (e) $t = 3/\omega$. Chiral edge motion is clearly visible.
  }\label{supfig:dynamics}
\end{figure*}

We now numerically demonstrate how the chiral edge mode in the $x$-$p_y$ plane can be observed through wave packet dynamics. We choose the strength of the artificial magnetic field to be $B = -\tan \theta = -1$ for the numerical simulation.
As an initial state, we consider three wave packets, located at the left edge, center, and the right edge of the system. To consider physics relevant to a single band or band gap, we project the wave packet onto the lowest level of each tube. Experimentally, in ultracold gases, such an initial state can be prepared by creating a wave packet out of single-particle ground states of each tube by starting from Bose-Einstein condensates in each tube. In photonics, where light can be injected from the outside, we can insert light with frequency corresponding to the lowest Landau level to prepare a similar initial condition. The wave function of the initial state is plotted in Fig.~\ref{supfig:dynamics}(a) and Fig.~\ref{supfig:dynamics}(b), in both the $x$-$y$ and $x$-$p_y$ planes. We then let the wave packet evolve freely. The wave functions in the $x$-$p_y$ plane after time $t = 1/\omega$, $t = 2/\omega$, and $t = 3/\omega$ are plotted in Figs.~\ref{supfig:dynamics}(c)-\ref{supfig:dynamics}(e). We observe that the wave packet at the left edge goes upward, whereas the wave packet at the right edge goes downward, showing that the edge state goes clockwise in a chiral manner. This motion of the wave packet at the edge is consistent with the artificial magnetic field in the $x$-$p_y$ plane; the strength of the magnetic field is $B = -\tan \theta < 0$, which implies that the cyclotron motion inside the bulk of the system is counterclockwise, and the edge state goes clockwise. We note that the absolute value of the wave function plotted in the $x$-$y$ plane looks identical during time evolution. This is because our initial state is projected onto the lowest Landau level, and thus the wave function in each tube is an eigenstate of the Hamiltonian for each tube, and thus the only time evolution appears in the phase of the wave function. The evolution of the phase in the $x$-$y$ plane appears as chiral edge motion in the $x$-$p_y$ plane.

We should also comment on the direction of the edge mode with respect to the energy spectrum, see Fig.~\ref{fig:energynado}(a). One sees that, in the energy spectrum, the energy goes up at large values of $n$. These states at large values of $n > 0$ are localized at the right edge. Since the energy goes up, one might want to conclude that the group velocity is positive at the right edge. However, as one sees from Fig.~\ref{supfig:dynamics}, this is not the case; the group velocity is negative at the right edge. This apparent discrepancy is because one needs to consider $-n$ as the conjugate variable of $p_y$. Namely, the horizontal axis of Fig.~\ref{fig:energynado}(a) should be flipped when one wants to find the correct group velocity. By flipping the horizontal axis of Fig.~2(a), one obtains the correct group velocity, which is negative at the right edge.

\section{Quantized Hall response}
Since Landau levels are topological, the Hall response should be quantized, which can be a signature of nontrivial topology obtained from the bulk property of the system. The topological Chern number $\mathcal{C}$ of Landau levels with $B < 0$ is one, $\mathcal{C} = 1$; we now investigate if this unit Chern number can be observed in our setup.

One way to see the topological nature of the system is through Laughlin charge pumping~\cite{Laughlin:1981}; on a two-dimensional cylinder with a perpendicular magnetic field, when a magnetic flux quantum is inserted through the cylinder, the state on the cylinder moves along the cylinder by one unit.
As we naturally have a cylindrical structure in the $x$-$p_y$ plane, it is natural to ask how Laughlin charge pumping appears in our setup.
As we see below, quantized pumping in our system occurs naturally due to construction.
A magnetic flux can be inserted through the cylinder in the $x$-$p_y$ plane by shifting the center of the harmonic trapping potential in the $x$-$y$ plane.
If we shift the central position of the harmonic trapping potentials by $\delta x$, the harmonic trapping potential acts as $\frac{1}{2}m\omega^2 (x-nx_0 -\delta x)^2$ for a given tube labeled by an integer $n$. In the $x$-$p_y$ plane, this shift corresponds to the shift of the magnetic vector potential to $\mathbf{A}(x,p_y) = (0, -x\tan\theta + \delta x \tan \theta)$. The additional contribution of $\delta x \tan \theta$ is a constant term in the magnetic vector potential, and thus if we integrate it over the $p_y$ direction, we see that the cylinder in the $x$-$p_y$ plane is penetrated by a magnetic flux $2\pi \delta x \tan \theta/y_0 = 2\pi \delta x / x_0$. Therefore, inserting a magnetic flux quantum of $2\pi$ corresponds to taking $\delta x = x_0$. This shift of $\delta x = x_0$ corresponds to no change in the overall Hamiltonian except that a tube previously labeled by an integer $n$ now acts as a tube with a label $n+1$. Thus, as in the Laughlin pumping, the overall Hamiltonian comes back to the original one after the shift of $\delta x = x_0$. In the $x$-$y$ plane, the shift of $\delta x = x_0$ corresponds to moving the entire system by an amount $\delta x = x_0$ to the right, thus any state should also move horizontally by $x_0$. This implies that any state on the cylinder of the $x$-$p_y$ plane should also move along the cylinder by $x_0$, which is the manifestation of the Laughlin charge pumping in our setup. 
Another way to see the topological nature of the system is through dynamical response of the system. In ultracold gases, quantized Hall response can be observed through the center-of-mass response upon application of a force (synthetic electric field) $\mathbf{E} = (E_x, E_{p_y})$~\cite{Dauphin:2013, Aidelsburger:2015, Price:2016}. When a synthetic electric field $E_i$ is applied along the $i$ direction, the center-of-mass velocity is~\cite{Price:2016}
$v^{\mathrm{C.M.}}_i = -\epsilon_{ij} \frac{2\pi}{A_{\mathrm{BZ}}}E_j \mathcal{C}$,
where $A_{\mathrm{BZ}}$ is the area of the Brillouin zone and $\epsilon_{x p_y} = -\epsilon_{p_y x} = 1$.
For our setup of a uniform magnetic field in the $x$-$p_y$ plane, there is no translational symmetry in the $x$ direction, and therefore we need to consider a magnetic unit cell and the magnetic Brillouin zone. A magnetic unit cell is an arbitrary rectangle in the $x$-$p_y$ plane which encloses one magnetic flux quantum. Since the strength of the artificial magnetic field is $|B| = \tan \theta$, the area of a magnetic unit cell is $2\pi/\tan \theta$. Then the area of the magnetic Brillouin zone is $(2\pi)^2$ divided by the area of a magnetic unit cell and thus $A_\mathrm{BZ} = 2\pi \tan \theta$. Therefore, the center-of-mass velocity should follow
\begin{align}
	v^{\mathrm{C.M.}}_i = -\epsilon_{ij} E_j \mathcal{C}/\tan \theta = \epsilon_{ij} E_j \mathcal{C}/B. \label{halleffect}
\end{align}
We should then be able to determine the Chern number of the Landau levels in the $x$-$p_y$ plane by monitoring the center-of-mass motion. As we see below, the quantized response upon adding $E_{p_y}$ is nothing but the Laughlin charge pumping we saw above and turns out to be trivially satisfied, whereas the response upon $E_x$ can provide observable signatures of the quantized Hall response.

We first consider the Hall response upon adding $E_{p_y}$, which can be applied by making the artificial magnetic vector potential time dependent $\mathbf{A} = (0, -x\tan \theta -E_{p_y} t)$.
Such a time dependence can again be introduced by changing the position of the minima of the harmonic trapping potentials. Changing the location of the trap minima from $ n x_0$ to $n x_0 + \delta x$, in a $y$-independent manner, the magnetic vector potential in the $x$-$p_y$ plane becomes $\mathbf{A} = (0, -x\tan \theta + \delta x \tan \theta)$ as before. This implies that the desired magnetic vector potential can be created by moving $\delta x$ as $\delta x = -E_{p_y} t / \tan \theta$.
Now, the quantized Hall response through the center-of-mass velocity has a clear physical meaning. When the motion of the trapping potential is slow enough, the center of mass should follow the motion of the trap minima $\delta x (t)$. Therefore, the center-of-mass velocity in the $x$ direction is $v^{\mathrm{C.M.}}_x = \partial_t \delta x (t) = -E_{p_y}/\tan \theta$; from Eq.~(\ref{halleffect}), this implies the correct relation $\mathcal{C} = 1$.

Adding $E_x$, the center-of-mass velocity follows $v^{\mathrm{C.M.}}_{p_y} = E_x \mathcal{C}/\tan \theta$, which can provide a more nontrivial observable signature of topology. A synthetic electric field $E_x$ can be applied by adding a linear potential gradient $-E_x x$ to the Hamiltonian. We numerically simulate the time evolution of a wave packet located at the center, projected onto the lowest Landau level, under the field $E_x$. In Fig.~\ref{fig:energynado}(b), we plot the center-of-mass position of the wave packet in the $p_y$ direction under time evolution for different values of $B = -\tan \theta$. We also plot the theoretical prediction from Eq.~(\ref{halleffect}) assuming $\mathcal{C} = 1$, which agrees well with the numerical simulation of $\langle p_y\rangle$. The numerical simulation also exhibits oscillation around the theoretical line with oscillation period $2\pi/\omega$, which is a detailed structure of the cyclotron motion. Thus, the quantized Hall response can be observed from the center-of-mass motion in the $p_y$ direction under the application of $E_x$, and the Chern number can be estimated from its slope.

\section{Interaction and vortices}
We now turn to interacting many-body cases.
An important feature is that a typical short-range interaction in the $x$-$y$ plane translates into long-ranged and unconventional interaction in the $p_y$ direction.
In particular, the formation of vortices, which is a characteristic feature of weakly interacting bosons in a magnetic field~\cite{Madison:2000, Cooper:2008, Fetter:2009, Spielman:2009, Lin:2009}, cannot be observed in the $x$-$p_y$ plane only with a contact interaction in the $x$-$y$ plane.
Nonetheless, we find that, by adopting a long-range interaction in the $y$ direction, vortices can still form in our setup as we discuss below.
We now focus on bosons with the mean-field interaction and ask if vortices can form in $x$-$p_y$ plane. 
We use description in terms of the Gross-Pitaevskii equation with the condensate wave function $\Psi_n (x,t)$:
\begin{align}
	&i\partial_t \Psi_n (x,t) = H_\mathrm{yhop} + H_{\mathrm{int}} + \notag \\
	&\left\{ -\frac{\partial_x^2}{2m} + \frac{1}{2}m\omega^2 (x - n x_0)^2
	+ \frac{1}{2}m\tilde{\omega}^2 x^2\right\} \Psi_n (x,t), \label{gpeq}
	\\
	&H_\mathrm{yhop} = J_y (\Psi_{n-1}(x,t) + \Psi_{n+1}(x,t)), \\
	&H_\mathrm{int} =  g_s |\Psi_n (x,t)|^2 \Psi_n (x,t) \notag \\ &\hspace{1cm} + g_1 (|\Psi_{n-1} (x,t)|^2 + |\Psi_{n+1} (x,t)|^2) \Psi_n (x,t).
\end{align}
We additionally included an overall weak harmonic potential with the oscillator frequency $\tilde{\omega}$ which confines the particles. We take $\tilde{\omega}$ to be much smaller than $\omega$ so that its only effect is to prevent particles from flying away.
We also added a small intertube hopping term, $H_\mathrm{yhop}$, which allows an exchange of particles between different tubes. This hopping term allows the system to establish an overall phase coherence. 
In the $x$-$p_y$ plane, such an intertube hopping term enters as a weak periodic potential of the form $2J_y \cos (y_0 p_y)$.
In the numerical calculations below, we take $\tilde{a}_\mathrm{osc} \equiv 1/\sqrt{m\tilde{\omega}} = 10 a_\mathrm{osc}$ and $J_y = -0.01\omega$.
The interaction term $H_\mathrm{int}$ contains the contact interaction with strength $g_s$ and also a nearest-neighbor (long-range) interaction between tubes with strength $g_1$.

Although the interaction $H_\mathrm{int}$ looks innocuous in the $x$-$y$ plane, it translates into rather unusual infinite and long-ranged interactions in the $x$-$p_y$ plane. Writing the Fourier transform of $\Psi_n(x)$ along the $y$ direction as $\Psi(x,p_y)$, and considering the corresponding second-quantized creation and annihilation operators $\hat{\Psi}^\dagger (x,p_y)$ and $\hat{\Psi}(x,p_y)$, the interaction term written in $x$-$p_y$ operators, up to an overall normalization factor, reads
\begin{align}
\sum_{p_1+p_2 = p_3 + p_4}&\left\{\frac{g_s}{2} + g_1  \cos \left(\frac{y_0(p_1 - p_2)}{2}\right) \cos \left(\frac{y_0(p_3 - p_4)}{2}\right) \right\} \notag \\
\int dx\, &\hat{\Psi}^\dagger (x,p_1) \hat{\Psi}^\dagger (x,p_2) \hat{\Psi} (x,p_3) \hat{\Psi} (x,p_4).
\end{align}
This is a {\it position conserving} interaction considering $p_y$ as a synthetic position coordinate.
The contact interaction $g_s$ is a uniform interaction in the $x$-$p_y$ plane as long as the position is conserved, and thus infinite-ranged.
On the other hand, the intertube interaction $g_1$ has a factor of $\cos (y_0\frac{p_1 - p_2}{2}) \cos (y_0\frac{p_3 - p_4}{2})$; the interaction is strongest when $p_1 \approx p_2$ and $p_3 \approx p_4$, which corresponds to the situation where the interacting particles are close to each other. Therefore, the intertube interaction introduces a shorter-ranged interaction along the $p_y$ direction compared with the contact interaction. We note, however, that, considering $p_y$ as a synthetic position, this intertube interaction is still an unconventional form which cannot be derived from a two-body potential depending only on the relative positions between two interacting particles.

\begin{figure}[t]
  \centering
  \includegraphics[width=\columnwidth]{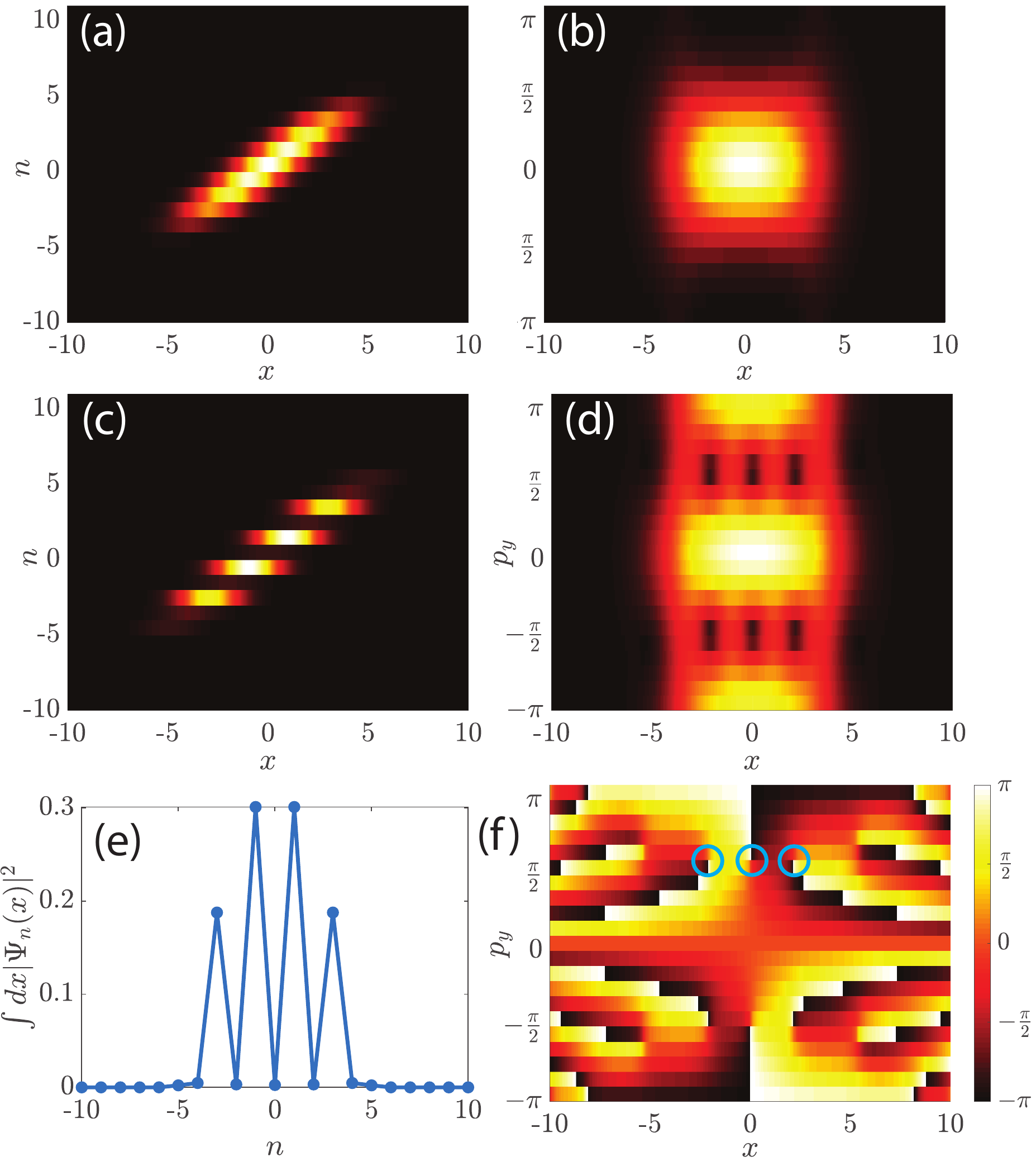}
  \caption{The interacting ground states. The ground-state wave function $|\Psi_n (x)|$ when $g_s =  a_{\mathrm{osc}} \omega $ and $g_1 = 0$, plotted in (a) the $x$-$y$ plane, and (b) the corresponding wave function in the $x$-$p_y$ plane, $|\Psi (x, p_y)|$. The ground-state wave function when $g_s =  a_{\mathrm{osc}} \omega$ and $g_1 = 2  a_{\mathrm{osc}} \omega$, plotted in (c) the $x$-$y$ plane, and (d) the corresponding wave function in the $x$-$p_y$ plane, where dips of the wave function can be observed. (e) The integrated density for each tube, $\int dx |\Psi_n (x)|^2$, plotted as a function of $n$; a density modulation is visible. (f) The phase profile of the wave function $\Psi (x,p_y)$. The phase singularities (vortices) at $p_y > 0$ are marked by blue circles. The units of length in the $x$ direction are $a_\mathrm{osc}$, and the units of $p_y$ are $1/y_0$.
  }\label{fig:interaction}
\end{figure}

We first consider the ground state when only the contact interaction $g_s$ is present.
In Figs.~\ref{fig:interaction}(a) and~\ref{fig:interaction}(b), we plot the ground state, obtained by imaginary-time propagation of the Gross-Pitaevskii equation (\ref{gpeq}), when $g_s = a_{\mathrm{osc}} \omega$ and $g_1 = 0$, assuming the normalization $\sum_n \int dx |\Psi_n (x,t)|^2 = 1$.
Figure~\ref{fig:interaction}(a) shows $|\Psi_n (x)|$ in the $x$-$y$ plane, whereas Fig.~\ref{fig:interaction}(b) shows $|\Psi (x,p_y)|$.
We see that the particles are clumped at the center of the $x$-$p_y$ plane, and no vortex-like structure is seen. This is because, in order to obtain vortices in the $x$-$p_y$ plane, we need a short-range repulsive interaction in $x$-$p_y$, but the interaction due to $g_s$ is infinite-ranged in the $p_y$ direction. 

To obtain vortices in the $x$-$p_y$ plane, we need to have a long-range interaction along the $y$ direction, which introduces a shorter-range interaction in the $x$-$p_y$ plane. 
Technology to realize long-range interactions in synthetic quantum matter has been rapidly developing; representative setups include atoms with dipole-dipole interactions~\cite{Lahaye:2009}, systems involving Rydberg states~\cite{Saffman:2010, Gorshkov:2011, Sevincli:2011}, and atoms in a cavity with light-mediated interactions~\cite{Ritsch:2013}.
Now we include a long-range interaction in the simplest form through the nearest-neighbor interaction $g_1$, which can be realized in ultracold atomic gases using dipole-dipole interactions with proper orientation of the dipole field~\cite{Lahaye:2009}.
In Figs.~\ref{fig:interaction}(c) and~\ref{fig:interaction}(d), we plot the ground state when $g_s =  a_{\mathrm{osc}} \omega$ and $g_1 = 2  a_{\mathrm{osc}} \omega$.  We see that $|\Psi (x,p_y)|$ has six dips. By looking at its phase profile in Fig.~\ref{fig:interaction}(f), we confirm that these dips are the vortices formed in the $x$-$p_y$ plane. In the $x$-$y$ plane, the state has density peaks at every other tube, as seen in the integrated density for each tube plotted in Fig.~\ref{fig:interaction}(e).
Vortices on a cylinder are known to have dynamical properties different from those on a plane~\cite{Ho:2015,Guenther:2017}; our proposal provides a unique setup where such phenomena can be studied.
Since the Hamiltonian has a discrete translational symmetry along the angle $\theta$ from the $x$ axis, the state breaks this translational symmetry, thus forming a supersolid state in the $x$-$y$ plane. Whether this supersolid structure persists in the absence of the confining potential $\tilde{\omega}$ in a thermodynamic limit is a question to be addressed in a future work.

We note a crucial difference between previous studies relating long-range interactions, such as a dipole-dipole interaction, and the formation of vortices in the presence of an artificial magnetic field~\cite{Lahaye:2009,Zhang:2016}. In these earlier works, the dipole-dipole interaction, an artificial magnetic field, and vortices all appear in the same (physical) two-dimensional plane. Our result, on the other hand, considers a long-range (intertube) interaction in the $x$-$y$ plane, but the vortices appear in the reciprocal $x$-$p_y$ plane where the artificial magnetic field is realized.

Below, we give a more detailed analysis of the vortices. We first show the wave function profile when a hypothetical contact interaction in the $x$-$p_y$ plane is present. We see that the vortices corresponds to a supersolid structure in the $x$-$y$ plane, confirming that the vortices we observed above are indeed created by the same mechanism as the ordinary vortices in the presence of a magnetic field. We then discuss how vortices form as one increases the nearest-neighbor interaction $g_1$.

\subsection{Vortices under a contact interaction in the $x$-$p_y$ plane}
\label{sec:vortex1}

\begin{figure}[htbp]
  \centering
  \includegraphics[width= \columnwidth]{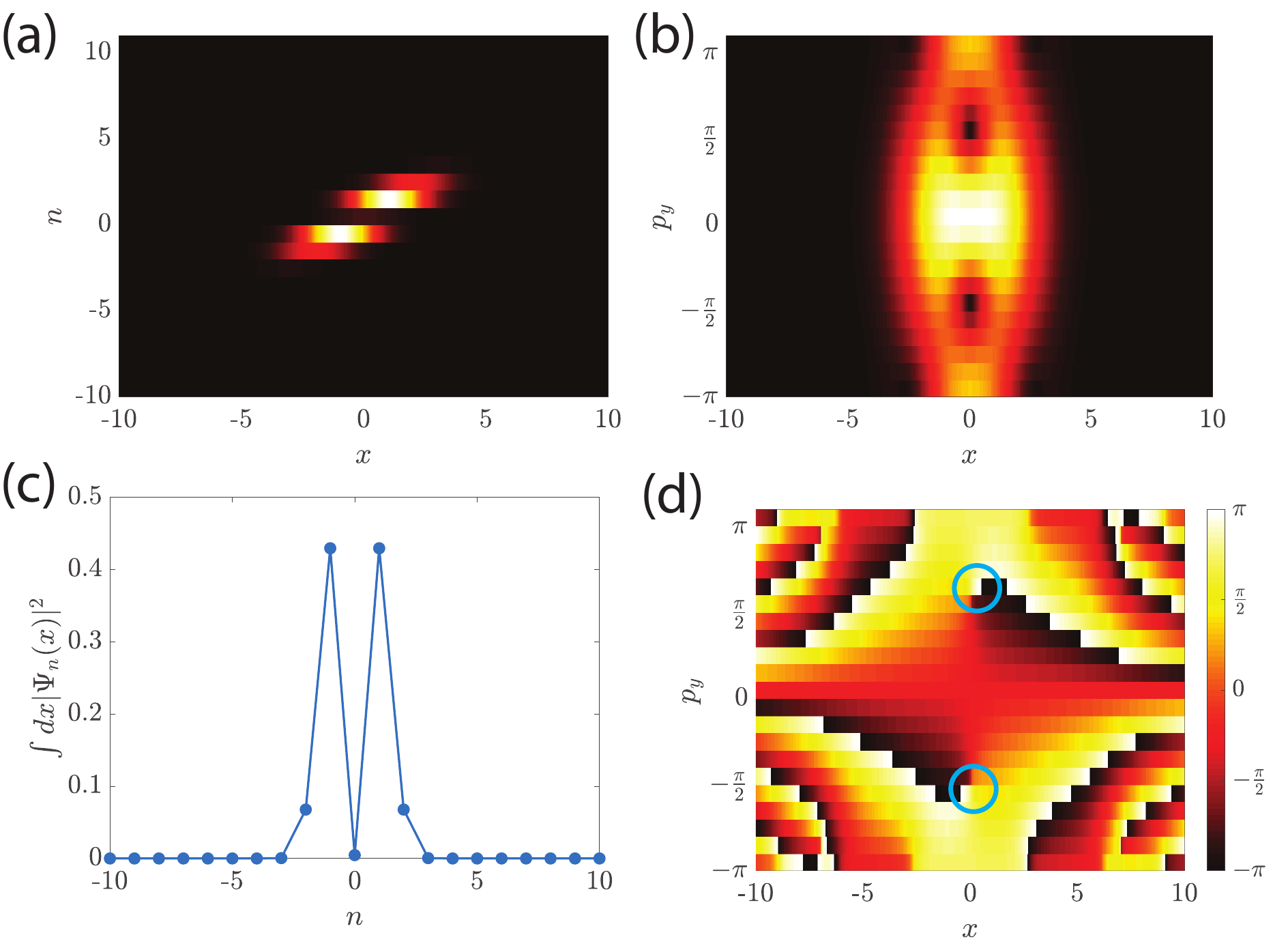}
  \caption{The mean-field ground state when we consider a hypothetical contact interaction in the $x$-$p_y$ plane. The units of length in the $x$ direction are $a_\mathrm{osc}$, and the units of $p_y$ are $1/y_0$. We take the strength of the contact interaction in the $x$-$p_y$ plane to be $\tilde{g}_s = 10 a_\mathrm{osc} y_0 \omega$, and no other interparticle interaction exists. Other parameters are $\tan \theta = 1$, $\tilde{\omega} = \omega/100$, and $J_y = -0.01 \omega$. (a) The absolute value of the wave function in the $x$-$y$ plane, where two density peaks are seen. (b) The wave function in the $x$-$p_y$ plane, where two dips of the wave function are seen. (c) The integrated wave function of each tube, $\int dx |\Psi_n(x)|^2$, from which one can see the density peaks more clearly. (d) The phase profile of the wave function in the $x$-$p_y$ plane; the two blue circles indicate the position of the phase singularities, namely the vortices.
  }\label{supfig:localxpy}
\end{figure}

We first analyze the bosonic ground state if the interaction is contact in the $x$-$p_y$ plane. Note that this is the situation one often encounters in the $x$-$y$ plane in ultracold atomic gases in the presence of an artificial magnetic field~\cite{Madison:2000, Cooper:2008, Fetter:2009}, and one expects the formation of vortices. We include the contact interaction in the $x$-$p_y$ plane by adding a term $\int dx \int dp_y \tilde{g}_s |\Psi (x,p_y)|^4$ in the energy functional, where $\Psi (x,p_y)$ is the normalized wave function in the $x$-$p_y$ plane. We choose $\tilde{g}_s = 10 a_\mathrm{osc} y_0 \omega$, $g_s = g_1 = 0$ and we take all the other parameters to be the same as before; in particular, we take $\tan \theta = 1$. In Fig.~\ref{supfig:localxpy}, we plot the ground state obtained by imaginary time evolution of the Gross-Pitaevskii equation. The absolute values of the wave function in the $x$-$y$ as well as the $x$-$p_y$ planes are plotted in Figs.~\ref{supfig:localxpy}(a) and~\ref{supfig:localxpy}(b), respectively. We can see two density peaks in the $x$-$y$ plane, similar to the structure seen in Fig.~\ref{fig:interaction} when the nearest-neighbor interaction is included. The total density of each tube is plotted in Fig.~\ref{supfig:localxpy}(c), where the vertical axis denotes the integral of $|\Psi_n (x)|^2$ over $x$. The corresponding wave function in the $x$-$p_y$ plane shows the vortex structure, as confirmed by the phase profile in Fig.~\ref{supfig:localxpy}(d). These vortices, formed by an artificial magnetic field and a contact interaction in the $x$-$p_y$ plane, are the ordinary vortices which have been observed in ultracold gases in the presence of an artificial magnetic field in real ($x$-$y$) space. Comparing these results with the wave function found in Fig.~\ref{fig:interaction}, we conclude that the vortex structure found in Fig.~\ref{fig:interaction} has the same origin as the conventional vortices in the presence of an artificial magnetic field.

\subsection{Vortex formation}
\label{sec:vortex2}

\begin{figure*}[htbp]
  \centering
  \includegraphics[width=1.8\columnwidth]{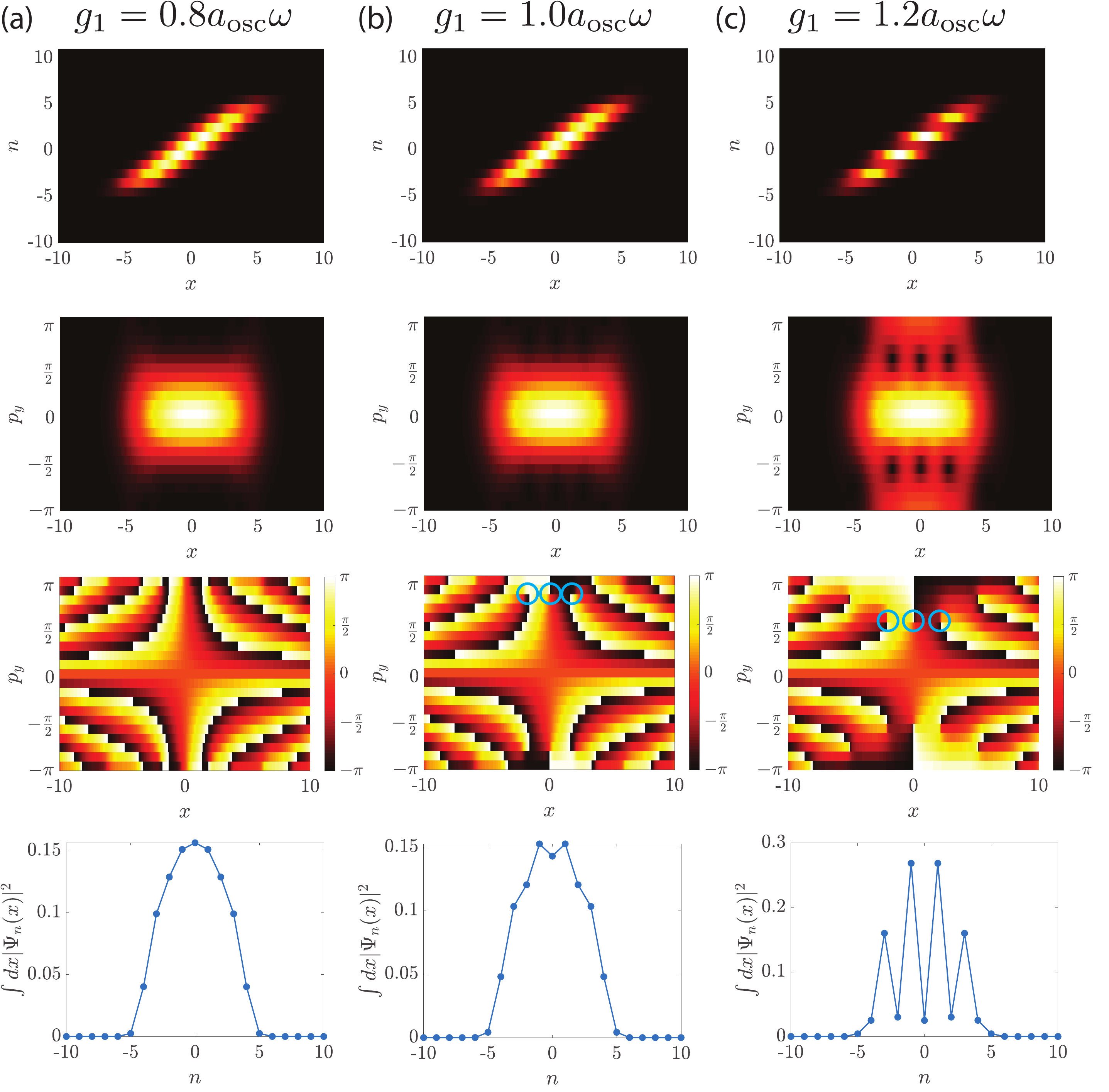}
  \caption{Formation of vortices in the $x$-$p_y$ plane. The three columns represent the ground states for $g_1 = 0.8 a_\mathrm{osc} \omega$, $g_1 = 1.0 a_\mathrm{osc} \omega$, and $g_1 = 1.2 a_\mathrm{osc} \omega$, respectively, from left to right. In each column, the top row shows $|\Psi_n (x)|$ of the ground state, the second row shows $|\Psi (x,p_y)|$, the third row shows the phase profile of $\Psi (x,p_y)$, and the bottom row shows $\int dx |\Psi_n (x)|^2$ as a function of $n$. In the phase profile panels, we marked phase singularities at $p_y > 0$ by blue circles. (There is also the same number of phase singularities at $p_y < 0$, which are not encircled.) The units of length in the $x$ direction are $a_\mathrm{osc}$, and the units of $p_y$ are $1/y_0$.
  }\label{supfig:vortex}
\end{figure*}

Now we come back to the interaction we originally considered: the contact interaction $g_s$ and the nearest-neighbor interaction $g_1$. We analyze how vortices form by varying the value of $g_1$ and looking at the wave function profile. 
In Fig.~\ref{supfig:vortex}, we plot, for three different values of $g_1$, the wave function in the $x$-$y$ plane and the $x$-$p_y$ plane, as well as the phase profile of the wave function in the $x$-$p_y$ plane and the integrated density for each tube.
We use $\tan \theta = 1$, $g_s = a_\mathrm{osc} \omega$, $\tilde{\omega} = \omega/100$, and $J_y = -0.01 \omega$.
We can observe that vortices form as one increases $g_1$.
We see that, when $g_1 = 0.8 a_\mathrm{osc} \omega$, there is no phase singularity in the $x$-$p_y$ plane, and the integrated density for each tube has a peak at $n = 0$ and monotonically decreases as one goes away from $n = 0$.
The situation changes for $g_1 = 1.0 a_\mathrm{osc} \omega$. Although it is still difficult to see vortex formation in $|\Psi (x,p_y)|$, by looking at the phase profile of $\Psi (x,p_y)$, we clearly see six phase singularities, indicating the formation of vortices. Correspondingly, density modulation starts to be seen in the integrated density for each tube.
When $g_1 = 1.2 a_\mathrm{osc} \omega$, the vortices become easily visible in $|\Psi (x,p_y)|$. The phase singularity and the density modulation also become more pronounced. 
The vortices are fragile against increase of the intertube hopping amplitude $J_y$.
If we use $J_y = -0.02 \omega$, the density dip in the $x$-$y$ plane, and hence the vortex structure in the $x$-$p_y$ plane, similar to the one found for the middle column of Fig.~\ref{supfig:vortex}, will not appear until we increase $g_1$ until around $g_1 \approx 1.3 a_\mathrm{osc} \omega$.
Thus, a stronger interaction is necessary to form vortices. This tendency can be understood in the following way:
Intertube hopping $J_y$ introduces a sinusoidal potential energy in the $x$-$p_y$ plane along the $p_y$ direction.
To create vortices in the $x$-$p_y$ plane, the interaction effect should be strong enough to overcome the sinusoidal potential created by $J_y$.
We note that the vortices can in principle form also when there is no intertube hopping, $J_y = 0$. In such a case, however, there is no reason for the entire system to develop a phase coherence, and thus the resulting structure in the $x$-$y$ plane is not a supersolid anymore; it would rather be a phase incoherent density wave.

\section{Experimental platforms}
Our proposal is relevant to various platforms of synthetic quantum matter.
In ultracold atomic gases, decoupled arrays of one-dimensional tubes can be realized with a two-dimensional optical lattice~\cite{Bloch:NP}. Our proposal can then be realized by superposing another harmonic trapping potential with an appropriate angle. The quantized Hall response and many-body physics are directly relevant in ultracold atomic gases~\cite{Dauphin:2013, Aidelsburger:2015, Price:2016}. 
Various photonic platforms are also relevant to our proposal. For example, exciton-polariton microcavity can be made into an array of one-dimensional wires~\cite{Boulier:2020,Amo:rv1}, and an additional harmonic potential can be applied by varying the width of each wire. Since it is extremely challenging to break the physical time-reversal symmetry in exciton-polaritons to open a topological gap~\cite{Klembt:2018}, our proposal provides a valuable alternative to explore Chern insulator physics in exciton-polariton microcavities.
Photonic cavity arrays and propagating waveguide geometry are also suited to realize our proposal, where harmonic potentials can be applied by varying the cavity or waveguide size along the $x$ direction~\cite{Hafezi:2013, Rechtsman:2013}; for these systems it is more natural to consider a lattice structure in the $x$ direction, which results in a coupled wire setup in the $x$-$p_y$ plane~\cite{Ozawa:2017}.
Edge-related physics are more easily accessible in such photon-related platforms~\cite{Hafezi:2013, Rechtsman:2013, Milicevic:2017}.

We note that our proposal depends crucially on the fact that properties in momentum space ($p_y$ direction) are observable in this synthetic quantum matter. In ultracold gases, the momentum-space wave function is accessible through time-of-flight imaging and, in photon-related systems, far-field imaging provides information on momentum space.

\section{Conclusion}
We have shown that a set of tubes with harmonic trapping potentials whose centers are shifted is equivalent to charged particles in a magnetic field in the $x$-$p_y$ plane.
The underlying idea of our proposal is to replace a momentum in one direction by position to realize a simpler Hamiltonian, which simulates the original Hamiltonian in a hybrid real-momentum space. This method is quite versatile; for example, with a similar mechanism, a collection of one-dimensional Aubry-Andr\'e model~\cite{Aubry:1980} (the Harper model) with shifted modulation phases simulates a two-dimensional lattice with a magnetic field, the Harper-Hofstadter model~\cite{Harper:1955, Hofstadter:1976}, in real-momentum space, without breaking the physical time-reversal symmetry.
Our proposal provides a simple and powerful method applicable to realize various Hamiltonians, opening a possibility for quantum simulation in synthetic quantum matter.

\begin{acknowledgements}
I am grateful to Bryce Gadway, the discussion with whom led to the initial idea of this paper. I also thank Iacopo Carusotto, Jean Dalibard, Nathan Goldman, Maciej Lewenstein, and Hannah M. Price for useful discussions.
I am supported by JSPS KAKENHI Grant Number JP20H01845, JST PRESTO Grant Number JPMJPR19L2, JST CREST Grant Number JPMJCR19T1, RIKEN Incentive Research Project, and the Interdisciplinary Theoretical and Mathematical Sciences Program (iTHEMS) at RIKEN.
\end{acknowledgements}

\end{document}